# From Cooperative Scans to Predictive Buffer Management

Michał Świtakowski
Actian Corporation
michal.switakowski@actian.com

Peter Boncz
CWI Amsterdam
p.boncz@cwi.nl

Marcin Zukowski
Actian Corporation
marcin.zukowski@actian.com

## ABSTRACT

In analytical applications, database systems often need to sustain workloads with multiple concurrent scans hitting the same table. The Cooperative Scans (CScans) framework, which introduces an Active Buffer Manager (ABM) component into the database architecture, has been the most effective and elaborate response to this problem, and was initially developed in the X100 research prototype. We now report on the the experiences of integrating Cooperative Scans into its industrial-strength successor, the Vectorwise database product. During this implementation we invented a simpler optimization of concurrent scan buffer management, called Predictive Buffer Management (PBM). PBM is based on the observation that in a workload with long-running scans, the buffer manager has quite a bit of information on the workload in the immediate future, such that an approximation of the ideal OPT algorithm becomes feasible. In the evaluation on both synthetic benchmarks as well as a TPC-H throughput run we compare the benefits of naive buffer management (LRU) versus CScans, PBM and OPT; showing that PBM achieves benefits close to Cooperative Scans, while incurring much lower architectural impact.

## 1. INTRODUCTION

Analytical databases systems are the cornerstone of many business intelligence architectures. Workloads in these systems tend to be characterized by large-scale data access: whereas transactional workloads zoom in on just a few records, analytical workloads often touch a significant fraction of the tuples. Consequently, full table scans and range-scans are frequent in such workloads. The frequency of scans has been increased further by the market trend to push away from the reliance on complex materialized view and indexing strategies, towards less DBA tuning-intensive systems with more predictable performance, often using *columnar storage*.

Consequently, a workload of *concurrent* analytical queries often consist of *concurrent* table scans. Traditionally, database systems employed simple LRU or MRU buffer management strategies, which causes concurrent scans to compete for disk access. This not only increases the latency of individual queries, as they take turns on the I/O resources, but also decreases the overall system I/O throughput, as different scans reduce the access locality and might cause thrashing.

In response to this problem, *circular scans* [5] were introduced in Microsoft SQLServer and other products, such that concurrent queries attach to an already ongoing scan, limiting thrashing. Subsequently, in DB2 the buffer manager was additionally changed to *throttle* queries [13] where a fast query that scans together with a slower query gets slowed down such that both stay together and keep sharing I/O. The most elaborate proposal in this direction came from the X100 research prototype[1] in the form of *Cooperative Scans* [21]. Cooperative Scans transform the normal buffer manager into an *Active Buffer Manager* (ABM), in which scans at the start of the execution register their data interest. With this knowledge of all concurrent scans, ABM adaptively chooses which page to load next and to pass to which scan(s), without having to adhere to the physical table order, always trying to keep as many queries busy as possible. To do this, ABM uses a flexible set of *relevance functions* which both strive to optimize overall system throughput and average query latency.

**Contributions.** The first contribution of this paper is to report on experiences in implementing Cooperative Scans in Vectorwise [18], a modern analytical database product using columnar storage, that evolved from the X100 research prototype in which the original Cooperative Scans were conceived. In the first versions of Vectorwise, Cooperative Scans were not supported, as integrating ABM turned out to be complex, in particular due to its interaction with concurrent updates, data reorganization and parallel query execution. We came round to implementing this now, and it turned out to be complicated yet feasible. In the course of the project, however, we also came up with an alternative approach that achieves much of the same goals, but with significantly less impact to the other system components. Thus, the second contribution of the paper is this new approach to concurrent scan buffer management, named *Predictive Buffer Management* (PBM). Rather that delivering data out-of-order as Cooperative Scans do, the main idea of PBM is to improve the buffer management policy. PBM tracks the progress of all scan queries, and uses this progress tracking to estimate the *time of next consumption* of each disk page. PBM exploits this knowledge to give those pages that are needed soonest a higher temperature such that they will be likely kept in the buffer. As such, this approach comes close to the perfect-oracle OPT algorithm [1]. While OPT is usually of theoretical interest only, as advance knowledge of all accesses is unrealistic, PBM exploits the fact that with long-running scan queries the system does have a good picture of the immediate future accesses. Our evaluation, both on the synthetic benchmarks from [21] as well as the throughput run of TPC-H, shows that for almost all buffer sizes as well as many disk speeds, PBM achieves benefits close to Cooperative Scans.

---

[1]The X100 system later evolved into Vectorwise – see www.actian.com/vectorwise

**Outline.** In Section 2 we re-cap the main ideas behind Cooperative Scans, before describing the technical challenges we had to solve in order to integrate ABM in a production system (Vectorwise) rather than a research prototype (X100). We then describe Predictive Buffer Management (PBM) as a simpler alternative in Section 3. In Section 4 we evaluate all approaches and also compare them to the theoretical optimal OPT algorithm, both on the synthetic benchmarks of [21] and the TPC-H throughput experiment. We finally describe future steps and related work in resp. Sections 5, and 6 before concluding in Section 7.

## 2. MATURING COOPERATIVE SCANS

**CScans Re-Cap.** Figures 1 and 2 contrast traditional buffer management with the Cooperative Scans (CScans) framework [21]. Whereas the loading decisions in the traditional approach are made by each scan operator individually, and the buffer manager is a cache managed with an algorithm like LRU, in Cooperative Scans loading decisions are taken by an Active Buffer Manager (ABM). The CScan operator is a scan that may produce data out-of-order, in the order it receives it from ABM. This flexibility is exploited by ABM to optimize both average query latency of all concurrent CScans as well as overall system throughput. ABM requires all CScan operations in the system to register their needed data ranges upfront with the RegisterCScan() function. Data is divided into large portions (*chunks*). Once a CScan is registered, it repeatedly asks ABM to deliver data to process (GetChunk()). This process is repeated until all data ranges registered by the CScan have been delivered. Finally, the CScan can unregister itself from ABM (UnregisterCScan()).

ABM executes in a separate thread that is responsible for the following operations:

- choosing a CScan for which it will deliver data.
- choosing a chunk that will be loaded.
- performing actual loading of the data from disk.
- waking up any blocked CScans that are interested in processing the loaded chunk.
- evicting chunks when the buffer space gets full.

To make the above-mentioned decisions, ABM uses four *relevance functions*. They assign priority to a CScan or a chunk depending on a specific operation that needs to be done. In particular: QueryRelevance() is computed on all active CScans, to choose which most urgently needs data. It aims to prioritize *starved* queries and short queries. Starved queries are queries that still need to process data, but have (almost) no available data in the buffer pool.

Once ABM has chosen which CScan operator to serve, the LoadRelevance() function is computed on all chunks to determine which data is most worthwhile to load for it. To maximize buffer reuse, it favors chunks that many other CScans are interested in.

In case a given CScan has more than one chunk available in the memory, ABM computes the UseRelevance() function for all cached chunks to decide which one to give to the CScan to process. To make chunks ready for eviction earlier, it chooses chunks that fewest CScans are interested in.

Finally, the decision which chunk to evict is made by computing KeepRelevance() on all cached chunks, and evicting the lowest scoring one (if it scores lower than the highest computed LoadRelevance()). This function aims to evict chunks that fewest CScans are interested in.

ABM makes decisions on a large granularity of a *chunk* (at least a few hundreds of thousands of tuples), rather than disk pages. One reason to choose such a granularity is to preserve sequential locality on the page level, even though the concurrent CScans will trigger scattered chunk load requests. The second reason is that with large chunks, there are many fewer chunks than there are pages, and ABM can use more computationally-expensive algorithms in its scheduling policy than for normal buffer management.

In column stores, it is wrong to think of a chunk as a set of pages. Rather, chunks have to be viewed logically as ranges of tuples; for instance each 100,000 consecutive tuples could be viewed as one chunk. The reason is that in columnar databases each column occupies a highly different amount of pages, due to different data types, and data compression ratio. One column from the same table might be stored on a single page, while other columns from that same table may take thousands of pages. When ABM decides to load (or evict) a chunk, in a column store this range of tuples will be translated to a set of pages, for each column involved. It may be the case that one page contains data from multiple adjacent chunks. This property of columnar storage is certainly a complication, but the general principles behind Cooperative Scans remain unchanged [21].

**Implementation Experiences.** We now discuss the most important issues that needed to be solved to fully implement CScan and integrate ABM in the Vectorwise system. Vectorwise is an analytical database system [12], created by integrating the upper layers (APIs, SQL parser, optimizer) of Ingres with the *vectorized* query execution and hybrid row- and columnar storage layers, which evolved from the X100 prototype [19].

The design of Cooperative Scans implies that ABM is a global object governing the behavior of all scan operators running in the system, yet in a full-fledged database system, introducing objects that have a global state has many complications. In the X100 [20, 21] research prototype, ABM and CScan were implemented without regards for updates, parallelism, or coexistence with traditional scan (which will still be used in situations where the query plan absolutely needs tuple order and hence cannot use a CScan).

We now discuss how we met the challenges that arose from implementing these missing features.

### 2.1 Updates

Vectorwise uses special in-memory differential structures called Positional Delta Trees (PDTs) [11] to handle modifications of a database. Differential updates avoid I/O when

1760

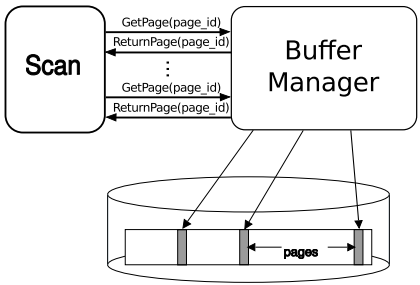
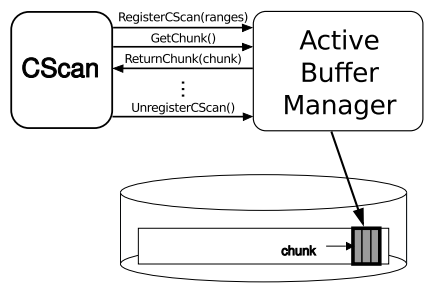
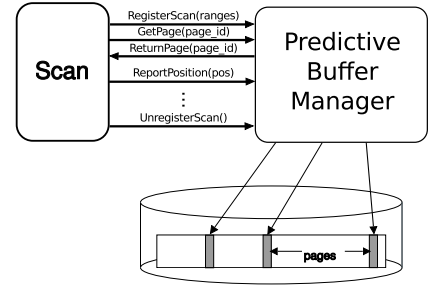

Figure 1: Traditional Buffer Manager    Figure 2: Cooperative Scans    Figure 3: Predictive Buffer Manager

updating columnar data (which would otherwise perform one I/O write per column per modified tuple!) and also avoid recompression CPU overhead, since columnar data would have to be decompressed, modified, and then compressed again. Each scan (classical `Scan` or `CScan`) operator therefore reads stale columnar data, but performs on-the-fly *PDT merging* to produce output that corresponds to the latest database state. PDTs organize differences *by position* rather than by key value, so merging is cheap CPU-wise and also avoids the need to read the key columns from disk.

Vectorwise implements snapshot isolation [2] to handle concurrent transactions, and it uses the PDTs to represent these snapshots. Rather than giving each transaction a separate copy of one big PDT structure, Vectorwise exploits the fact that differential structures can be stacked (differences on differences, etc.), so a snapshot consists of the stable table on disk, together with three layers of PDTs. Only the topmost and smallest PDT is private to a particular snapshot, the other PDT levels are shared between transactions [11]. Therefore, the cost of snapshot isolation in terms of memory consumption is low.

**PDT Merge in `CScan`.** Our first challenge was to allow `CScan` operators to support PDT merging. The main difference between the `Scan` and `CScan` operators is that the `Scan` operator receives data in-order, whereas the `CScan` operator receives data out-of-order in a chunk-at-time manner. This complicates the PDT merging operation: applying updates by position is easier if all tuples arrive in positional order, than when they (periodically) jump around due to the out-of-order data delivery.

First, we explain PDT merging in general; later we come back at the complications due to the out-of-order delivery.

The updates stored in PDTs are indexed by position, and merging revolves around two kinds of tuple positions: SIDs and RIDs. The Stable ID (SID) is a 0-based dense sequence identifying tuples stored in the stable storage, i.e. the state of a table without updates applied. The Row ID (RID) is a 0-based dense sequence enumerating the stream that is visible to the query processing layer, i.e. after updates are applied to data fetched from the stable storage. The SID and RID are different unless all PDTs are empty, i.e. PDT merging is an identity operation.

The PDT is a tree data structure that stores Delete, Insert and Modification actions, organized using SID as a key. Each PDT node also contains the *running delta*, which is the difference between RID and SID caused by the updates in the underlying PDT subtree. The PDT data structure is stored in RAM and has logarithmic CPU cost for updates, as well for RID-SID translation – this translation can be performed in both directions.

RIDs are very volatile under updates (a delete in the middle of the table decreases all subsequent RIDs) and are therefore not stored anywhere, but generated during PDT merging. SIDs follow implicitly from the physical tuple order on disk and are not stored.

For tuples belonging to the stable storage (stable tuples) not marked as deleted in PDTs, the RID of that tuple can be translated to its SID, as every stable tuple has a unique SID. For tuples that are not part of the stable storage, i.e. inserts stored in the PDT, their RID translates to the SID of the first stable tuple that follows it. As a consequence, for new tuples inserted in PDTs only, there may be multiple tuples that are assigned the same value of SID. Thus, it is not possible to define SID to RID conversion as an inverse of RID to SID conversion, because the RID to SID conversion is not an injective function. However, it is possible to introduce two possible variants of SID to RID conversion. A certain SID can be translated either into lowest RID that maps to it, or the highest one. We introduce the `RIDtoSID()` function to perform RID to SID conversion, and `SIDtoRIDlow()` and `SIDtoRIDhigh()` functions to perform the „low" and „high" variants of SID to RID conversion, respectively.

Figure 4 depicts an example of conversion between SID and RID. In the lower part, we can see stable tuples stored on the hard disk and their respective SIDs. After tuples are loaded and merged with changes stored in PDTs, we obtain a new stream of tuples that is enumerated with RID. Deleted tuples are marked with red, whereas the inserted ones with green. Arrows indicate the translation between SID and RID. The blue arrows indicate results of the `SIDtoRIDlow()` function, whereas the red ones correspond to `SIDtoRIDhigh()`. Note that some of the arrows are only one-way. They indicate tuples, for which it is not possible to reverse the RID to SID conversion, because they are in the middle of a sequence of tuples with the same value of SID. The deleted tuples are stored on the hard disk and loaded into memory but get eliminated by the PDT merging process. Thus, there is no RID that translates into the SID of those tuples. However, it is still possible to translate their SIDs to RIDs. The assigned RID is the lowest RID that translates into a SID higher than the one of the deleted tuple.

The Active Buffer Manager has been designed to work purely on the storage level. It is not aware of the existence of in-memory updates and PDTs in particular – concurrent queries may execute in different snapshots and have different PDT data. Thus, the concept of ABM chunks as logical tuple ranges is defined in terms of SIDs. However, the `CScan` range-scan operator is initialized in the query plan with a list of RID ranges that it has to produce for its parent operators. Those RID ranges must hence be converted into SID ranges that ABM operates on.

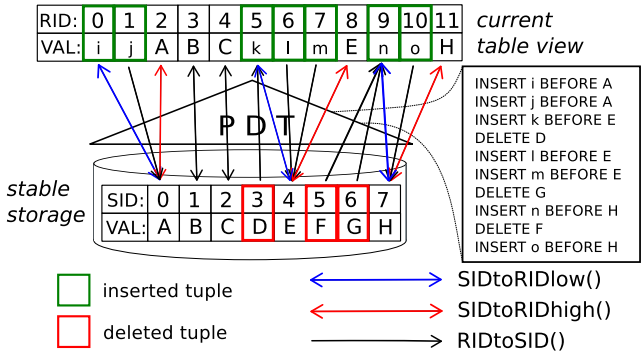

Figure 4: Conversion between SID and RID

Once ABM loads a chunk and passes it to the `CScan` operator, an inverse process has to be performed. As the RID to SID translation is not reversible, the algorithm is more complicated. Given a certain chunk, which is effectively a range of stable tuples that are ready in the memory, we must determine what RID range can be generated out of them. The lower SID boundary is translated with `SIDtoRIDlow()`, and the upper boundary with `SIDtoRIDhigh()` to get widest possible RID range that can be generated. Note that the SID range of neighboring chunks after translation to RID using the same approach may *overlap* with the RID range of chunk that has been retrieved. Thus, it is necessary to keep track of all RID ranges that have already been processed by the `CScan` operator. This is the major new requirement coming from the fact that `CScans` receive data out-of-order from ABM. Once a new SID range (i.e. chunk) is delivered by ABM and translated to RID range, it needs to be trimmed to make sure no tuples are generated twice. Also, whenever a new chunk starts, the PDT merging algorithm needs to be re-initialized. This involves finding a proper position in the PDT structure, where the process of merging should start.

**Bulk Appends.** For bulk inserts, Vectorwise avoids filling PDTs with huge amounts of insert nodes, but rather allows physical data appends to the stable tables. For such `Append` operators, the system must also provide a snapshot isolation mechanism. This type of snapshot is a memory-resident set of arrays that contain references to pages (page identifiers) belonging to each column of the table (one array per column). Adding new data is performed by creating new pages and adding references to them in the snapshot.

ABM needs to be aware of these storage-level snapshots. Additionally, since it is responsible for loading the pages, it has to determine how tuple ranges registered by `CScan` operators translate to pages that need to be loaded. This becomes more complex when *concurrent* queries appending data are executed. Figure 5 presents four simple transactions that add new data using the `Append` operator and scan the table afterwards with a `CScan`. Transaction T2 differs from the others by using the `Commit()` statement to make its changes persistent. Suppose T3 and T4 start after T2 ends and T1 starts before T2's end. An example course of execution of those transactions and snapshots they are working on is presented in Figure 6. Snapshot 1 represents the initial state of the table. The table consists of four pages identified with numbers from 0 to 3. It is the master snapshot, that all transactions start with, thus it is marked with red. T1 and T2 begin with appending new data to the table. As a result, two different transaction-local (marked with blue) snapshots

```
T1 Append(table, <data stream 1>)
T2 Append(table, <data stream 2>)
T1 CScan(table)
T2 CScan(table)
T2 Commit()
T3 Append(table, <data stream 3>)
T4 Append(table, <data stream 4>)
T3 CScan(table)
T4 CScan(table)
```

Figure 5: Schedule for Transactions T1-T4

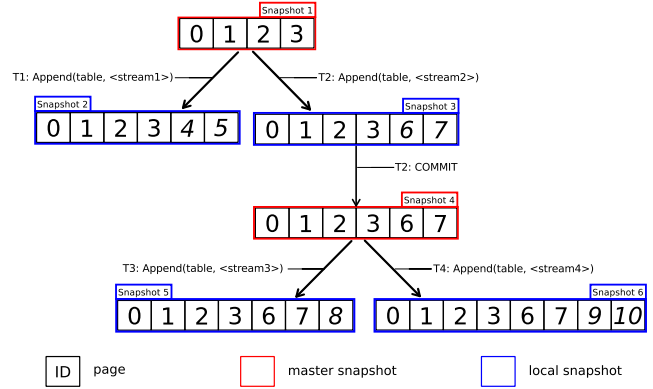

Figure 6: Snapshot Isolation for Appends in Vectorwise

are created: Snapshot 2 and Snapshot 3. They share references to the first four pages but their last two pages are different, as they were created separately for each transaction by `Append` operators. T2 commits its applied changes while T1 does not. Consequently, the snapshot that T2 worked on becomes the master snapshot (marked with red). From now on, all new transactions will use Snapshot 3. Thus, T3 and T4 append data to the new snapshot.

The crucial observation is that all transactions running in the system work on snapshots that have a common prefix. For example, the common prefix consists of pages {0, 1, 2, 3} when T1 and T2 are the only transactions working in the system.

In the presented scenario ABM needs to detect transactions working on different snapshots. The fact that snapshots have common prefixes can be exploited, so that queries can benefit from sharing opportunities regardless of differences in snapshots. To achieve that, we extend ABM with the notions of *shared* chunks and *local* chunks. Shared chunks consist of pages that belong to at least two snapshots that are used by transactions at certain moment. Pages of local chunks belong to only one snapshot. It should be noted that chunks encompass multiple columns. Thus, a chunk can be regarded as shared only if all its pages in all columns belong to snapshots of two transactions. Even after appending a single value to a table, its last chunk becomes local.

Figure 6 can be also interpreted on the chunk level instead of the page level. Let us focus on such a situation from the perspective of ABM. Shared chunks can be detected by finding the longest prefix that belongs to at least two snapshots. Suppose that transactions T1, T3 and T4 are working in the system, so the longest shared prefix is the set {0, 1, 2, 3, 6, 7}. Consequently, we can exploit that knowledge and load those chunks earlier as it increases sharing opportunities.

Note that the longest shared prefix can change as queries enter and leave the system. For example, when T1 and T2

are the only working transactions, the shared prefix consists of chunks {0, 1, 2, 3}. However, when T2, T3 and T4 work in the system, the longest shared prefix that can be found is the set {0, 1, 2, 3, 6, 7}. An important observation is that the longest shared prefix is in fact the only shared prefix that can be found. Suppose that at certain moment there are two transactions whose snapshots contain pages {0, 1, 2, 3, 4, 5} (Snapshot 2 in Figure 6) and two transactions working on pages {0, 1, 2, 3, 6, 7} (Snapshot 3 in Figure 6). Having two transactions working on identical snapshots means that this snapshot must have been the master snapshot at the moment these transactions started, or one transaction already committed (making its snapshot the master snapshot) before the other started. Thus, both Snapshot 2 and Snapshot 3 must have been master snapshots at some moment. This is impossible, because they have a non-empty intersection, but none of them is a subset of another, i.e. none of them could have been created by appending to another. In Vectorwise, only one of the concurrent transactions that applied Appends to its snapshot can commit its changes and make its snapshot the master snapshot. The other transactions are detected to be conflicting and consequently aborted. Thus, there may be concurrent transactions working either on Snapshot 2 or Snapshot 3, but not on both.

Every time a new CScan scanning a certain table enters the system, ABM finds the longest prefix that is shared by at least two CScan operators and marks chunks accordingly. Chunks belonging to the found prefix are marked as shared, whereas the other ones are marked as local. The same procedure is performed when a CScan leaves the system as it may influence the longest shared prefix as well. Keeping track of shared and local chunks allows to maximize sharing opportunities in case transactions work on similar, but not equal versions of the same table. Moreover, it allows proper adjustment of metadata used for scheduling. In particular, shared chunks have higher chances to be loaded and kept in the memory longer, whereas local chunks are loaded and used only once, typically in the final phase of a CScan.

**PDT Checkpoints.** PDT checkpointing [11] is an aspect closely related to PDT updates. When updates are applied to the database, the memory usage of PDTs may grow considerably, and this can not be sustained forever. At some point, the contents of the PDTs have to be migrated to disk, creating a new stable table image. A simple implementation of this process involves scanning the table from the stable storage, merging updates stored in PDTs and storing the result as a new version of the same table. After checkpointing finishes, all incoming queries will work on this new table image. Note that for the new version of the table a new set of pages is allocated, i.e. the old and the new version do not share any pages. As depicted in Figure 7, a checkpoint results in creating a new master snapshot (marked as red) that consists of new pages. After the checkpoint is finished and the new version of the master snapshot is created, all starting transactions will use Snapshot 2. At the same time, there may be still transactions in the system working on the old Snapshot 1. Once all transactions using Snapshot 1 are finished, Snapshot 1 can be destroyed.

ABM needs to detect that a certain table has been checkpointed, as transactions using Snapshot 1 and Snapshot 2 work on non-overlapping sets of pages, thus cannot share pages loaded in the buffer pool. Every time a new CScan is registered, ABM verifies the set of pages belonging to the snapshot of transaction that CScan is a part of.

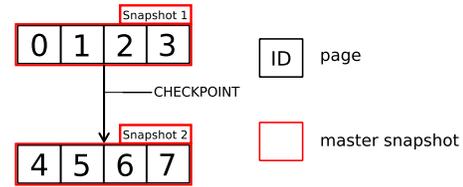

Figure 7: Behavior of a checkpoint. PDT changes are applied to create a new snapshot.

There are four possible cases to handle:

(i) It is the first CScan that accesses that particular table – ABM has to create and initialize metadata for chunks, i.e. register new chunks.

(ii) There are other CScans working on the same table and their snapshot is identical – ABM does not need to change its metadata related to tables.

(iii) There are other CScans working on the same table and a common prefix with their snapshots can be found – ABM has to find shared and local chunks as discussed before.

(iv) There are other CScans working on the same table and the snapshot of the new CScan contains different pages that the other snapshot – ABM registers new version of the same table along with its chunks (as in (i)).

The metadata in ABM needs to reflect changes in the database after checkpoint was finished. Thus, every time a CScan operator finishes and un-registers itself, ABM verifies whether there are other CScans working on a snapshot that is identical or has a common prefix with the snapshot of the leaving CScan. In case there are no such CScans the metadata related to chunks belonging to that version of the table is destroyed.

## 2.2 Parallelism

Vectorwise employs intra-query parallelism that is implemented by means of a set of Exchange (XChg) operators [9]. A query plan of a parallelized query is transformed to contain multiple copies of certain subtrees that are connected with special XChg operators. Those subtrees are executed in parallel by multiple threads. An example of such a transformation is depicted in Figure 8.

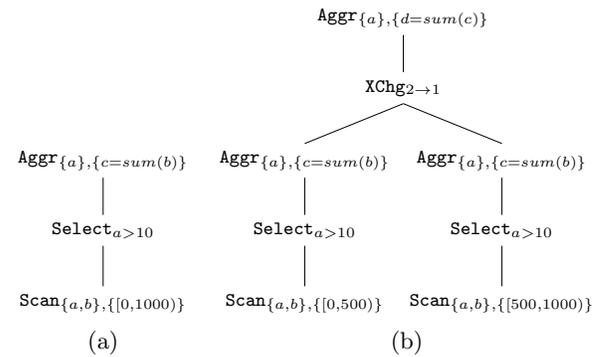

Figure 8: An example of a transformation of a sequential query plan (a) into a parallel query plan (b)

The subtree under the `Aggr` operator has been duplicated and connected with the `XChg` operator that merges two streams into one. On top, there is another `Aggr` operator added that computes the overall aggregate result. In particular, the `Scan` operator has been replaced by two separate `Scans` that each scan a separate part of the original RID range. In Vectorwise, RID ranges are split equally between working threads and assigned to each of the `Scans` as presented in Equation 1.

$$\text{range } [a..b] = \begin{cases} \text{range } [a .. a + \frac{(b-a)*1}{n}) \\ \text{range } [a + \frac{(b-a)*1}{n} .. a + \frac{(b-a)*2}{n}) \\ \vdots \\ \text{range } [a + \frac{(b-a)*(n-1)}{n} .. a + \frac{(b-a)*n}{n}) \end{cases} \quad (1)$$

The current implementation of parallelism for Cooperative Scans is functional but not yet optimal. In parallel plans, we statically partition one CScan into multiple RID ranges, creating one CScan for each thread; similar to the policy for traditional scans. Currently, ABM is not aware that those `CScan` operators belong to the same parallel query plan. In most cases, `CScans` belonging to the same parallel query plan request the same number of chunks and process them at similar pace. However, if one or more threads turns out to be much faster than the others, ABM will start treating one of `CScans` as shorter, thus prioritize it. Unfortunately, this will have negative impact on the execution time of a parallel query as it will create parallel imbalance. Secondly, the usage of `CScans` in parallel query plans may lead to increased data skew. Regardless of the actual range to be scanned, `CScan` operators process data in chunk-at-time manner and have to wait until full chunk is loaded, but the chunk range is trimmed to a common RID boundary, as described in Section 2.1. A `CScan` whose range encompasses few chunks has to load more data than really needed. This problem becomes more relevant after a single `CScan` is divided into multiple `CScans` scanning smaller ranges.

As an example of these sub-optimalities, suppose the parallelization generates a plan with two `CScan` operators scanning chunks from 0 to 4 and from 5 to 9 respectively. Let us assume that the buffer pool contains a set of the following chunks: {3,5,6,7,9}, so there is one chunk available for the first `CScan` but four for the second one. Thus, the second `CScan` will only have to wait for loading chunk 8, whereas the first one needs to load four chunks. Consequently, the first `CScan` has to wait for much more I/O, delaying the execution of the whole query. A better approach would be to distribute the already-loaded chunks evenly between participating `CScans` and also distribute evenly the newly loaded chunks. We defer this to future work.

## 2.3 Scan/CScan Coexistence

In a complex analytical DBMS there are scenarios where the approach of out-of-order chunk-at-time data delivery is suboptimal or cannot be used. Examples are queries that:

- exploit data that is stored in-order on disk. Storing a table as a clustered index in Vectorwise leads to a physical tuple ordering according to the index key [19], and this can e.g. lead to plans that exploit ordered aggregation, which cannot accept data out-of-order.

- have an access pattern that is much more fine-grained than a chunk. Vectorwise has automatic MinMax indexes [20] that may restrict the scan ranges significantly, and e.g., query plans with late column projection might generate scans that retrieve many small RID ranges.

In such situations, the normal `Scan` operator is used. In the current implementation of Cooperative Scans, ABM pins a fixed amount of pages in the buffer pool. The disadvantage of this simple approach is that we must statically determine this memory budget, dividing the available RAM resources between the two buffering schemes.

In our approach to integrating ABM and the traditional buffer manager, we ended up giving the full buffer pool to ABM, and extend it with an option such that a `CScan` can demand in-order delivery of chunks. In that case, flexibility is lost, but the `CScan` becomes a drop-in replacement for `Scan`. The consequence is that range scans now always use a large granularity (chunk). We realize that in systems built for a mixed workload this is a disadvantage.

We considered an alternative integration of ABM and the traditional buffer manager by creating a common page eviction policy; which means that we would need to create a common temperature with which chunks and pages could be compared. The common policy would need a new implementation of `KeepRelevance()` accepting both pages and chunks. One idea to do so is to estimate the *time of next consumption* of a page or chunk. This is possible to achieve in both cases. Estimating the time of next consumption for pages processed by a `Scan` operator involves estimating its speed and distance to the given page. In case of chunks it is more complicated, but still feasible. Every chunk that is cached in the memory can be assigned a score with `UseRelevance()`. Sorting chunks by the score allows to determine expected order of consumption. Finally, by combining it with a speed estimate of a certain `CScan`, the expected consumption time can be calculated. The eviction policy would try to evict data with highest expected time of consumption first. We did not implement this integration of Cooperative Scans and the traditional buffer manager yet, but the idea of estimating a time-of-next-consumption led to the creation of Predictive Buffer Management.

## 3. PREDICTIVE BUFFER MANAGEMENT

In literature on buffer management policies, the OPT algorithm [1] represents the perfect-oracle algorithm that provably [15] obtains best performance. OPT usually is only of theoretical importance, as its requirement is a perfect and full knowledge of the order of all future page accesses. Given that, the OPT algorithm simply evicts the page that will be referenced furthest in the future. When trying to find an integrated page temperature metric for the traditional buffer manager and ABM, we came upon the idea of estimating *time of next consumption* for pages that are the subject of a table scan. This can be done by measuring over time the speed with which each `Scan` moves through the data. This time of next consumption is a close estimate of the metric used in the OPT algorithm. We exploit the fact that in an analytical database system with long-running scans, it *is* possible to predict the near-term future accesses to a large extent.

The resulting Predictive Buffer Manager (PBM) is a novel buffer management policy that is well suited for analytical workloads with many large scans. It shares the same basic principle with Cooperative Scans – exploiting knowledge about currently working queries. In contrast, it does not introduce a global object where all decisions regarding loading and evicting pages are made. Similarly to traditional scan processing (see Figure 1 and 3), I/O requests are made by Scan operators. PBM is responsible for evicting pages when the buffer pool is full, and does *not* change the order of data access requests. As a result, PBM can be incorporated into a database system without influencing its higher layers. Moreover, it is independent of physical storage organization. ABM, on the other hand, needs separate implementation and relevance functions for NSM and DSM [21].

The interaction between PBM and Scan operators is done by means of three functions: RegisterScan(), UnregisterScan() and ReportScanPosition() as shown in Figure 3. As in the case of Cooperative Scans, RegisterScan() passes information about future page accesses of a starting Scan. To perform its predictions, PBM keeps track of the current position and speed of processing of each Scan. ReportScanPosition() is invoked periodically to update the position and processing speed of Scans. Finally, UnregisterScan() informs PBM that a Scan finished and its metadata can be freed.

Figure 9 presents pseudocode of the most important functions in PBM. Let us focus on RegisterScan() first. Its purpose is to iterate over all pages that are going to be requested by the Scan operator and register them. This involves finding the number of tuples that the Scan will need to process before a certain page is requested and saving it along with a scan ID in the page.consuming_scans collection. Finally, PagePush() is executed to recalculate the priority of a page. Priorities of pages are assigned by the PageNextConsumption() function, which calculates the estimated time when the page will be needed. Given a Scan that will need the page, this is a matter of dividing the current distance to that page in tuples by the observed speed of the Scan. PageNextConsumption() returns the nearest time (the minimum) that any of the Scan operators that registered the page will need it.

**PBM Data Structures.** Buffer management algorithms performed on the page level need to be CPU-efficient. PBM thus needs to provide an efficient way to perform operations such as registering a page, un-registering a page when it is consumed and changing its priority. One possible solution is using a binary heap or another priority queue structure that provides $O(log(n))$ complexity for most operations. However, we found this solution to incur too much overhead in Vectorwise, especially in highly-concurrent scenarios. We implemented an alternative approach that maintains low overhead and allows to benefit from our idea.

The crucial observation is that we do not strictly need to have a fully accurate priority queue. To amortize the cost of eviction, pages are evicted in groups of 16 or more pages. Thus, we only need to have an ability to retrieve a group of pages whose priorities are the lowest or almost the lowest. To enable that, PBM partitions pages into buckets depending on their expected time of consumption. In more detail, the PBM buckets are organized in a timeline, where each bucket represents a time range, and consecutive buckets are grouped in *bucket groups*. Within a bucket-group, all buckets have the same time range length; and this length increases exponentially for each further bucket group. At the end of the timeline there is a „not requested" bucket, which is used to store pages that are not registered by any scans i.e. page.consuming_scans is empty. The start of the timeline represents the current moment.

```
 1: function REGISTERSCAN(table, columns, range_list)
 2:     id ← GetNewScanIdentifier()
 3:     for all col in columns do
 4:         tuples_behind ← 0
 5:         for all range in range_list do
 6:             {get a collection of pages belonging
 7:             to specified column and range}
 8:             pages ← GetPages(col, range)
 9:             for all page in pages do
10:                 {register the scan id and the number of tuples
11:                 it will have to read before consuming this page}
12:                 page.consuming_scans ←
13:                     page.consuming_scans ∪ (id, tuples_behind)
14:                 tuples_behind ←
15:                     tuples_behind + page.tuple_count
16:                 {recalculate the priority of the page
17:                 and push to appropriate bucket}
18:                 PagePush(page)
19:     return id
20:
21: function PAGENEXTCONSUMPTION(page)
22:     nearest_consumption ← NULL
23:     for all (id, tuples_behind) in page.consuming_scans do
24:         scan ← GetScanById(id)
25:         next_consumption ←
26:             (tuples_behind − scan.tuples_consumed)/scan.speed
27:         if nearest_consumption = NULL or
28:             next_consumption < nearest_consumption then
29:             nearest_consumption ← next_consumption
30:     return nearest_consumption
31:
32: procedure PAGEPUSH(page)
33:     { if page belongs to a bucket, remove it from there}
34:     if page.bucket ≠ NULL then
35:         BucketRemovePage(page.bucket, page)
36:     nearest_consumption ← PageNextConsumption(page)
37:     if nearest_consumption = NULL then
38:         BucketAddPage(not_requested_bucket, page)
39:     else
40:         bucket_number ←
41:             TimeToBucketNumber(nearest_consumption)
42:         BucketAddPage(buckets[bucket_number], page)
43:
44: procedure REFRESHREQUESTEDBUCKETS
45:     for i ← 0 to Size(buckets) - 1 do
46:         if time_passed mod buckets[i].length = 0 then
47:             buckets[i − 1] ← buckets[i]
48:             buckets[i − 1].length ← buckets_length[i − 1]
49:     for i ← 0 to Size(buckets) - 1 do
50:         if buckets[i] = NULL then
51:             buckets[i] ← NewBucket()
52:     for all page in buckets[−1] do
53:         PagePush(page)
54:     BucketDestroy(buckets[−1])
55:
56: procedure EVICTPAGE
57:     if not Empty(not_requested_bucket) then
58:         chosen_bucket ← not_requested_bucket
59:     else
60:         for i ← Size(buckets)-1 downto 0 do
61:             if not Empty(buckets[i]) then
62:                 chosen_bucket ← buckets[i]
63:                 brake
64:     page ← BucketPopPage(bucket)
65:     Evict(page)
```

Figure 9: Implementation of the Predictive Buffer Manager

Figure 10 depicts some „requested buckets" and the relation between bucket number and time range. In general, PBM uses $n$ groups of $m$ buckets each, where the length of each range doubles in each successive group. Thus, to represent a time range $[0, max\_time]$ it is sufficient to create $\lceil \log_2(\frac{max\_time}{m}) \rceil$ groups. The first group represents the shortest range whose length equals *time_slice* constant. The following values are used in the example depicted in Figure 10: $n=m=2$, *time_slice*=100ms.

When the next consumption time of a page is calculated by PageNextConsumption(), it is placed in one of the above-

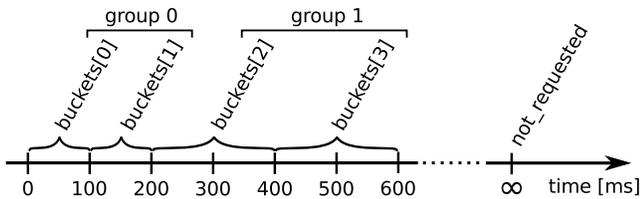
Figure 10: PBM Buckets

mentioned buckets, by the `PagePush()` function, that uses a static index accessed with the `TimeToBucketNumber()` function, so translating time to bucket number is an $O(1)$ operation. Buckets are implemented as doubly-linked lists, so add and remove operations are also $O(1)$.

The eviction algorithm implemented in the `EvictPage()` function first tries to evict pages belonging to the ,,not requested" bucket i.e. the pages that will not be accessed by currently working queries. If that is impossible, PBM starts evicting pages from requested buckets starting from the highest numbers i.e. the ones with the highest estimated time of next consumption.

The buckets holding requested pages become outdated as the time passes. Thus, after *time_slice* time passes, the numbers of buckets need to adjusted. This process is performed by `RefreshRequestedBuckets()`. It involves shifting buckets left depending on their length and current time. A certain bucket is moved by one position if the time that passed since it was last moved equals the length of the time range of the bucket. Since `RefreshRequestedBuckets()` is always called after *time_slice* time passes, it is sufficient to check if the *time_passed* variable used as a time counter is divisible by bucket length. Also, if a bucket is shifted to a different group, its length needs to be divided by 2. The first loop of `RefreshRequestedBuckets()` shifts buckets according to that rule and updates their lengths. Positions that were left empty are filled with new buckets in the subsequent loop. The first bucket is shifted to position -1 ($buckets[-1]$). If predictions were accurate it should already be empty. Otherwise, priorities of blocks belonging to it are recalculated and the bucket is destroyed.

**PBM/LRU.** PBM emulates the OPT algorithm quite closely for all paging decisions considering large tables that are being scanned. However, it does not have information on future queries. In our current implementation, therefore, we organize the pages in the last bucket (,,not requested") in LRU fashion as a doubly-linked list, as usual for LRU.

This simple solution, however, treats pages that are currently not of interest to a `Scan` as always having a lower priority than those who are. Note that e.g. pages that are very frequently used by short-running scans, would thus be penalized, at those times that these short-running queries do not run (i.e. most of the time). In analytical workloads, this will happen for small (dimension) tables: despite being frequently accessed, PBM will evict their pages and I/O will be needed for them, which could be avoided. The performance impact of this is not large, though, as small tables contribute little I/O to the overall workload.

The issue of reconciling PBM with small queries is more relevant for mixed workloads, where parts of the data are accessed in OLTP fashion, and parts in OLAP fashion. We now outline an idea to deal with such situations, creating an algorithm with two sets of *counter-rotating* buckets. The idea is that there would be two sets of bucket groups:

- the PBM buckets of registered scans as described before, which one might depict *above* the timeline, which move to the *left* as time passes by, and
- the LRU buckets containing pages not of interest to an active scan, which one might depict *below* the timeline, which move to the *right* (aging).

Thus, instead of moving pages which are no longer useful for any `Scan` to the last bucket as basic PBM does, this PBM/LRU algorithm moves them into the LRU bucket corresponding to an estimated time of next consumption, based on observed usage frequency over a history of multiple queries. This estimate could for instance be created by keeping for each page the timestamps of the last four uses, and taking the average distance between them.

Each *time_slice* interval, the PBM buckets are moved to the left as described before, and highly similarly, LRU buckets are moved right (aging). Thus, for cached pages without an active scan, PBM/LRU estimates the next-time-of-consumption based on past history and applies aging by gradually moving them to the right side, using a rationale similar to LRU.

The eviction algorithm starts from the last buckets of both the LRU and PBM sets, first evicting pages in the LRU bucket; and if more is needed evicting pages from the PBM bucket of the same time range; moving left in time iteratively, if more pages need to be evicted.

We leave implementation of this algorithm as future work, and remark that it would best be evaluated in a database system other than Vectorwise, targeted at such a mixed workloads and providing proper indexing structures (such as secondary indices) for those. Our evaluation inside Vectorwise purely focuses on analytical workloads.

## 4. EVALUATION

We evaluate the various concurrent scan approaches on the same set of synthetic microbenchmarks used when the original Cooperative Scans were proposed [21], but supplement these with a *throughput run* on TPC-H 30GB as generated by its `qgen` tool, which subjects the system to concurrent query streams.

Our Linux test server is equipped with two 4-core Intel Xeon X5560 processors, 48 GB or RAM and 16 Intel X-25 SSDs in software RAID-0 on five SATA controllers, delivering well over 2GB/s of sequential bandwidth. This represents the upper end of what is possible in a local attached I/O subsystem, and in our experience typical customer configurations tend to be provisioned much lower. To simulate slower I/O configurations, in certain experiments we artificially slow down the bandwidth by limiting the rate of delivering pages from the storage layer of Vectorwise to the buffer manager. This allows us to test the algorithms on an I/O subsystem that provides bandwidth in a range from 200MB/s to 2GB/s. In all benchmarks we maximally exploit CPU power by enabling intra-query parallelism (see Section 2.2) with maximum number of threads per query set to 8.

In our experiments we compare traditional buffer management (LRU) with Cooperative Scans (CScans), Predictive Buffer Management (PBM) and OPT. To simulate the OPT algorithm, we gathered a trace of all page references that were made in the PBM run. We chose PBM as it is a scan order-preserving policy, and is designed to generate an I/O sequence closest to a scenario where we could precisely predict order and time of all page references. We then run an OPT simulator on this trace, and report the I/O volume caused by this. It should be stressed that in a dynamic environment every factor has influence on the final result. In particular, the trace would be different if we generated it against a policy having knowledge of future queries. The comparison with OPT is intended to get an idea of the optimal I/O volume of order-preserving policies (PBM, LRU) and to compare this with non-order-preserving policies (CScans).

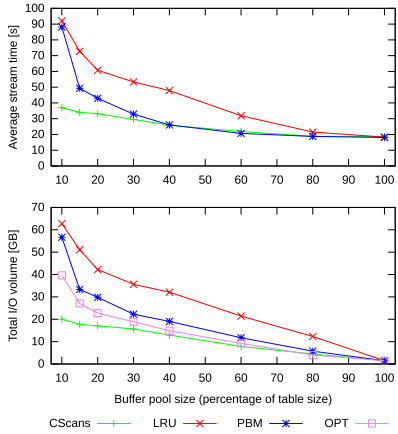

Figure 11: Microbenchmark results, varying the buffer pool size

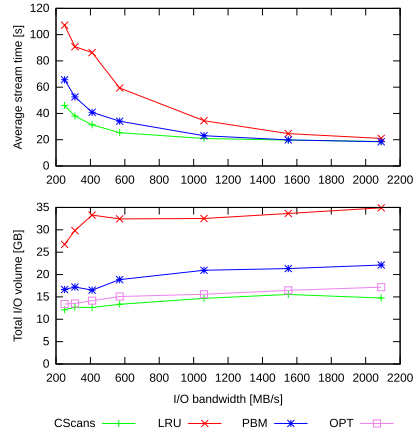

Figure 12: Microbenchmark results, varying the I/O bandwidth

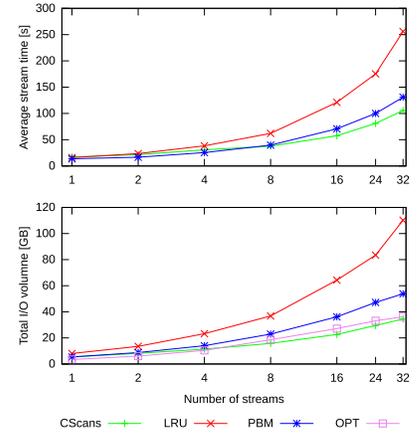

Figure 13: Microbenchmark results, varying the number of streams

### 4.1 Microbenchmarks

The microbenchmarks similar to those used in [21] are based on TPC-H queries Q1 and Q6, that scan the largest table (lineitem), performing selection, projection and aggregation. We parametrize each query with a tuple range that starts at a random position. The percentage of table size that is scanned in one range-scan is at chosen from the set {1%, 10%, 50%, 100%}. We run between 1 and 32 concurrent streams of query batches consisting of 16 queries. We explore three dimensions: size of the buffer pool, number of concurrent streams and I/O hardware bandwidth. As performance measure, the average stream time and total volume of performed I/O are used. Unless stated otherwise, in all benchmarks the following default parameters are used: 8 concurrent streams, buffer pool capacity equal to 40% of accessed data volume and I/O bandwidth of 700 MB/s.

**Buffer Pool Size.** The efficiency of a buffer management policy becomes more important as the volume of accessed data grows. We therefore vary the the buffer pool size to simulate growing datasets, from 10% to 100% of the total volume of data accessed by queries (ca.1550MB).

Figure 11 depicts the results. Lowering the size of the buffer pool below 100% immediately results in a higher I/O volume. The growth of I/O volume is distinctly highest for LRU. In case of LRU, the user-perceived performance starts to deteriorate significantly with buffer pool sizes lower than 80% of the data size, as the system is CPU-bound up to this point. PBM and CScans provide nearly constant performance for buffer pools holding 40% of data and higher. For very low buffer pool sizes, PBM and CScans start to diverge. Especially for a 10% buffer pool, PBM performance drops to the level of LRU. Cooperative Scans perform well even on very small buffer pools, because they have an ability to synchronize scans, which results in maximizing buffer reuse. This is not possible for PBM, which ends up with many scans scattered across the table. As a consequence, PBM has no possibility to increase buffer reuse.

The results of OPT place in between CScans and PBM. Since OPT does not change the order of page references, its performance deteriorates for very small buffer pools just like PBM. The fact that the order-preserving OPT is clearly beaten when the order-preserving assumption is dropped (CScans), is food for thought about proven optimality vs. implicit assumptions.[2]

**I/O Bandwidth.** We now check how the speed of the underlying I/O subsystem influences overall performance. Figure 12 depicts system performance measured by average stream time and total I/O volume. It is clear that I/O bandwidth strongly influences execution time. The difference between PBM and CScans disappears if bandwidth exceeds about 1GB/s, while the difference between LRU and others roughly disappears for 2GB/s and more. This is the point where the system becomes CPU bound. However, the total I/O volume remains approximately constant for all policies. It turns out that lowering the rate of page delivery slows down the query execution in a way that causes in-order page references only. Consequently, similar decisions are made by all scan-optimizing buffer management policies.

**Number of Streams.** Increasing the number of concurrent streams creates an opportunity to reuse the same page many times. However, it also results in more I/O requests being executed, possibly more scattered across the table, which makes it harder to exploit this opportunity. To make streams more homogeneous, we use a mix of Q01 and Q06 queries, all scanning 50% percent of the table starting at a random position. Figure 13 depicts benchmark results with the number of streams varying from 1 to 32. Both average stream time and total I/O increase in similar fashion with growing number of concurrent streams. Both PBM and CScans cope much better in experiments using more than 4 streams, confirming their usefulness in highly concurrent workloads.

---

[2]If we would use the CScans I/O trace to run OPT, its performance would have been equal or better than CScans.

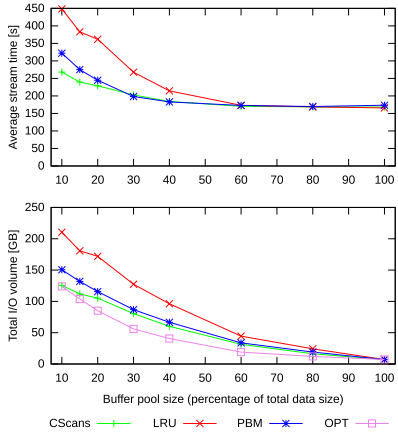
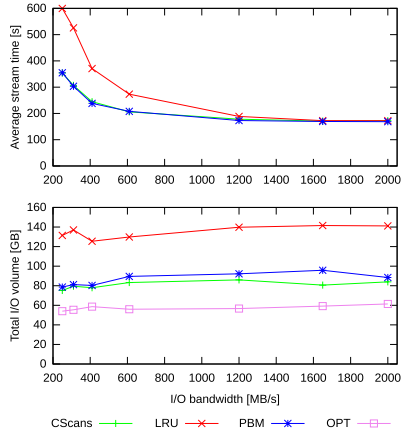
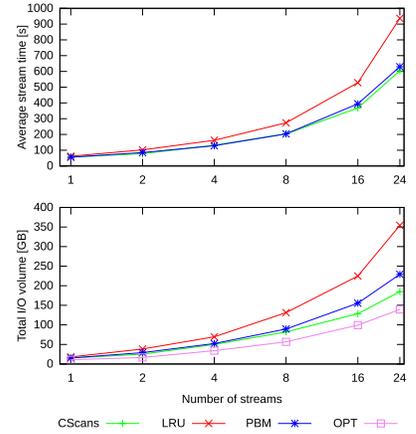

Figure 14: TPC-H throughput results, varying the buffer pool size

Figure 15: TPC-H throughput results, varying the I/O bandwidth

Figure 16: TPC-H throughput results, varying the number of streams

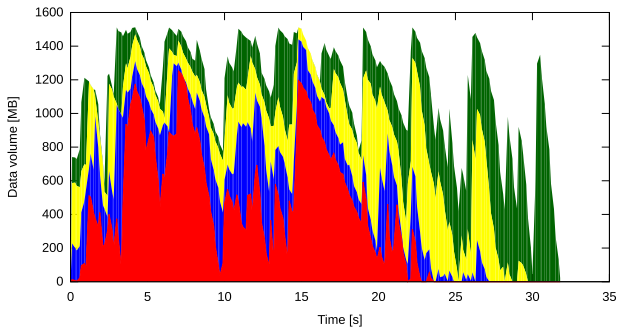
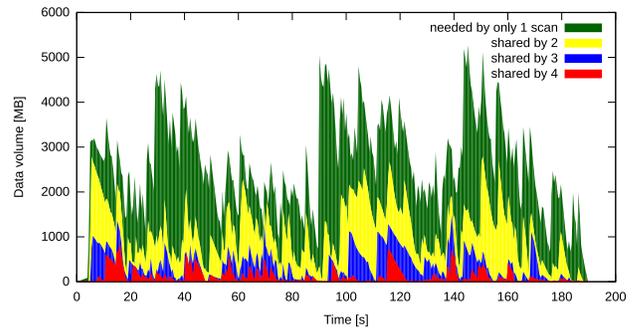

Figure 17: Sharing potential in microbenchmarks

Figure 18: Sharing potential in TPC-H throughput

## 4.2 TPC-H

The microbenchmarks only access a single table (lineitem). To evaluate the performance in more complex scenarios, we use the TPC-H throughput benchmark. The TPC-H experiments use the same 30GB scale factor as in the microbenchmarks.[3] It should be noted that that the TPC-H throughput run is much more complex than the microbenchmarks described in Section 4.1. The TPC-H schema consists of eight tables with 61 columns in total. The query set contains 22 queries with a high degree of complexity, typically more CPU intensive than the ones from the microbenchmarks. As a result, in the graphs we choose to depict results with a default I/O bandwidth of 600MB/s, and a default buffer pool size of 2250MB, which is 30% of the total volume of data accessed by all queries with 8 streams (ca.7500MB).

**Buffer Pool Size.** Figure 14 shows that a smaller buffer pool results in a higher I/O volume and execution time. The system is I/O bound for buffer pools smaller than 60% of data volume. With larger buffer pools, CPU work is the bottleneck, as the average stream time remains constant regardless of the buffer size. Two differences in comparison to Figure 11 can be observed. First, the gap between CScans and PBM is lower: PBM achieves performance similar to CScans for all buffer pool sizes. Secondly, there is no strong degradation for PBM in case of a 10% buffer pool.

**I/O Bandwidth.** As presented in Figure 15, lowering the I/O bandwidth in TPC-H experiments results in higher ex-ecution time, whereas the volume of I/O remains constant. Again, in case of TPC-H we observe the performance of PBM and CScans to be almost equal. The difference between all buffer management policies is eliminated when the I/O subsystem delivers 1200MB/s or more. This proves that the TPC-H workload becomes CPU-bound in more cases than the microbenchmarks described in Section 4.1.

[3]These results are reported for academic interest only, have not been audited and are not official TPC-H scores.

**Number of Streams.** Figure 15 confirms the advantage of PBM and CScans to be increasing for workloads with more concurrent streams. Maximally 24 streams could be executed on our server with 48 GB or RAM. Despite the fact that the streams were not homogeneous (i.e. used different queries scanning different parts of different tables) a pattern similar to the experiment in Figure 13 is observed.

**Sharing Potential.** To understand the performance results better, we provide an analysis of the *sharing potential*. In a system loaded with concurrently working queries, at any moment in time, one can count for each page how many active scans still want to consume it. Thus, one can compute the volume of data that is needed at some moment by only one scan, exactly two scans etc.

The result of this analysis is depicted in Figures 17 and 18. The green area indicates the amount of data that is needed by a single scan, thus is unlikely to be reusable by another query. The area marked with red represents the data that is needed by 4 or more scans working in the system at some moment. We see a clear difference between the microbenchmarks versus the TPC-H throughput run: the former has much more reuse potential than the latter. During the exe-

1768

cution of the microbenchmark we find large volumes of data useful to two or more Scan operators. On the other hand, in the TPC-H experiment, green area dominates the graph i.e. most of the data is going to be requested and processed by only one scan.

We conclude that the performance benefit that PBM and CScans achieve over simple policies like LRU strongly depends on the sharing potential.

**Overall Trends.** In the microbenchmark, which was designed to provide many I/O sharing opportunities, we see that both CScans and PBM provide significant benefits. In TPC-H there are still significant differences between LRU on the one hand and CScans and PBM, though the smaller reuse opportunities in this workload make the advantage of CScans over PBM really small. The only situation where PBM is clearly inferior to CScans, is the combination of extreme memory pressure (10% of the working set or less) and many reuse opportunities (microbenchmarks).

In the OPT simulation we observe a consistent trend. The results of OPT tend to be worse than results of Cooperative Scans in microbenchmarks, whereas they are better than Cooperative Scans in TPC-H. This observation can be explained by the higher complexity and lower sharing potential in TPC-H. Indeed, a more complex workload is less predictable, thus it is harder for CScans to make the right decisions based on the knowledge of the active queries. At the same time, the lower sharing potential reduces opportunities of out-of-order delivery used by CScans.

## 5. DISCUSSION

The paper started reporting on our project to make Cooperative Scans a full feature of the Vectorwise analytical database system. We think that we identified all major challenges and ways to address them. Further steps to fully complete the code have been mentioned in Section 2 as:

- a tighter integration with multi-core parallelism such that chunks can get assigned dynamically to different threads, rather than using static range partitioning.
- the development of a tight integration between traditional buffer management and CScans, so that queries could still use the traditional Scan operator, if the chunk granularity is too coarse.

While working on this latter point, we came upon Predictive Buffer Management, which turned out to be a much simpler solution to the problem of concurrent scans than CScans; that achieves almost the same benefits.

When considering the adoption of a new technique in a production system, not only the benefits it provides in the intended use case, but all its interaction with the rest of the system in all use cases, as well as issues like software complexity and reliability weigh in. As such, the choice to adopt PBM instead of CScans was a clear-cut decision, and it will become available in Vectorwise version 2.5.

As for the current state of PBM, we identified in Section 3 the following topics for future work:

- a more elaborate integration of buffer management across multiple queries (LRU) with the PBM framework. Currently, pages that will be requested by an active scan are prioritized over other pages. The sketched PBM/LRU algorithm addresses this issue, but needs further experimentation.

- improvements in PBM to address its weakpoint: extreme memory pressure in a workload with a lot of sharing potential (see Section 4), without compromising the simple architectural characteristics of PBM.

Regarding the latter point, the benchmark results in Figure 11 suggest that the performance of the PBM drops when the systems is using a very small buffer pool with many queries scanning at different positions of the table. In such a scenario PBM is not able to exploit sharing opportunities because it needs to deliver data in-order. We now shortly discuss two possible directions for improvement:

**PBM Attach & Throttle.** We could introduce circular-scan techniques into PBM, by allowing an incoming Scan to ,,attach" to already running Scans to make them share loaded data. This could be enhanced further along the lines of [13, 14] by throttling some queries, so that groups of queries scanning at close positions are formed. The algorithm that forms groups of queries can be extended to take information used by the PBM into account. Suppose the PBM keeps track of the next consumption time of pages that were lastly evicted from the buffer pool. Let us denote it by next_consumption_evict. A page whose next consumption time is higher or close to next_consumption_evict is likely to be evicted without being reused. In PBM, every page that has just been consumed is assigned a new next consumption time. By comparing it to next_consumption_evict, we can detect whether the consumed page is likely to be evicted or reused again. A Scan can be throttled if this would lower the next consumption time of pages it recently consumed to a value below next_consumption_evict. This would allow Scans working behind this particular Scan to catch up and benefit from pages loaded for the throttled Scan.

**Opportunistic CScans.** The Cooperative Scans framework assumes that all loading and eviction decisions are strictly managed by the Active Buffer Manager on chunk-at-a-time basis. This assumption makes the implementation more difficult as ABM needs to manage the global system state as well as all transaction states to provide consistent data access. A simpler approach to benefit from out-of-order delivery would be if the Scan itself would dynamically change the area in a table that is processed by a certain query, depending on the actual state of the buffer pool, without centralized planning. The Scan operator could constantly monitor which parts of the scanned table contain most cached pages. When a certain region of a table that is cached to a large extent is detected, the Scan could dynamically change the range to scan that region and possibly increase the number of times those pages are reused before being evicted. That way, Scans may *automatically* ,,attach" and cooperate.

## 6. RELATED WORK

Disk scheduling policies have been studied at length in operating systems research [16]. The classic policies developed include First Come First Served, Shortest Seek Time First, SCAN, LOOK and others. Also, in the area of virtual memory [1] and file system caching policies there were several policies introduced including LRU and MRU. It should be noted that the developed policies were aimed at access patterns and workloads different from the ones found in an analytical DBMS, where we have more knowledge of the work-load and also computationally more expensive algorithms can be used for scheduling.

In a DBMS it is possible to identify several access patterns depending on the type of a query and used indices such as looping sequential, straight sequential and random [4]. Depending on the access pattern, a different buffer management policy may be used. In [4] and subsequent work [3,6] scans were considered trivial and handled by either LRU or MRU policy.

The idea of developing policies aimed at increasing data reuse in an environment with multiple concurrent scans was introduced in commercial DBMS systems including Teradata, RedBrick [7] and Microsoft SQLServer [5]. The proposed solutions used either the elevator or attach policy that were compared to Cooperative Scans in [21].

The idea of circular scans is exploited in systems that are designed to handle a highly concurrent workload with tens to hundreds of queries. The main purpose for using this technique is to maintain high performance of the system irrespectively of the number of concurrent queries. The Crescando system [8,17] introduces Clock Scan where one thread services all queries working in the system. The QPipe architecture [10] is another example of a system using a circular scan that feeds multiple operators with data.

Another approach for improving data reuse in a multi-scan environment was introduced in IBM DB2 system [13, 14]. The idea is to group scans with similar speed and position together to let them reuse loaded data. The groups are controlled by allowing throttling of faster queries or recreating the groups if a considerable desynchronization occurs. In principle, this approach changes the access pattern to increase locality of references, but does not change the order of processing, nor the buffer management policy.

Under the assumption that full information about all future page accesses is known, it is possible to formulate an optimal algorithm i.e. an algorithm that minimizes the number of pages that have to loaded from disk to the buffer. The OPT algorithm [1] (also called MIN, Clairvoyant or Belady's Algorithm) governs a buffer provided a sequence of all future references is available. A short proof of optimality of OPT can be found in [15]. The Predictive Buffer Manager that we present in Section 3 is based on the idea of approximating the OPT algorithm.

## 7. CONCLUSIONS

In this paper we reported on the experiences in turning the Cooperative Scans (CScans) framework [21] from a research prototype into a product feature. This posed a number of challenges, as the Active Buffer Manager (ABM), central to the CScans framework, manipulates global system state and interacts with many other components of the database system. Concretely, we had to work on properly handling transactional updates (both bulk and trickle), multi-core parallelism, and the integration of CScans with the traditional buffer manager. While all these challenges were met, CScans have not yet been fully integrated into a product. The reason was the invention of a much simpler extension to traditional buffer management, called Predictive Buffer Management (PBM). This new technique turned out to provide benefits similar to CScans, while requiring much less disruption to the system architecture. The main idea behind PBM is that, in analytical workloads, careful monitoring of active table scans gives the system an opportunity to predict the future page accesses. As such, PBM is an approximation of the optimal page replacement algorithm OPT [1], which exploits such, usually unattainable, advance knowledge.

We believe that PBM is a great addition to the family of database buffer management policies, and we have outlined various avenues of its future improvement. We also hope that the story of CScans inspires researchers in academia to evaluate innovations not only in terms of absolute performance benefits, but also other criteria, such as software complexity and intrusiveness to system architecture (less is more!).